\documentclass[aps,prl]{revtex4}

\begin{document}

\title{Remark on orbital precession due to central-force perturbations}

\author{ O.~I.~Chashchina}
\affiliation{Department of physics, Novosibirsk State University, 630 090, 
Novosibirsk, Russia }

\author{ Z.~K.~Silagadze}
\affiliation{Budker Institute of Nuclear Physics and 
Novosibirsk State University, 630 090, Novosibirsk, Russia }

\begin{abstract}
This is a comment on the recent paper by G.~S.~Adkins and J.~McDonnell
``Orbital precession due to central-force perturbations''  published in
Phys. Rev. D75 (2007), 082001 [arXiv:gr-qc/0702015]. We show that the main 
result of this paper, the formula for the precession of Keplerian orbits 
induced by central-force perturbations, can be obtained very simply by the 
use of Hamilton's vector. 
\end{abstract}

\maketitle

In the recent paper G.~S.~Adkins and J.~McDonnell reconsidered the old
problem of perihelion precession of Keplerian orbits under the influence
of arbitrary central-force perturbations. Their main result is the formula
for perihelion precession in the form of a one-dimensional integral 
convenient for numerical calculations. 

The reason why this well studied and essentially textbook problem \cite{2} 
came into the focus of the current research is the recent usage of this 
classical effect to constrain hypothetical modifications of Newtonian gravity
from higher dimensional models \cite{3}, as well as the density of dark 
matter in the solar system \cite{4}.

Traditionally the simplest way to study the perihelion motion is the use
of the Runge-Lenz vector \cite{5,6}. The Runge-Lenz vector
\begin{equation}
\vec{A}=\vec{v}\times\vec{L}-\alpha\,\vec{e}_r
\label{eq1}
\end{equation}
is the extra constant of motion originated from the hidden symmetry of the
Coulomb/Kepler problem \cite{7}. Here $\alpha=GmM$, $\vec{L}$ is the angular
momentum vector and $\vec{v}$ is the relative velocity of a planet of mass
$m$ with respect to the Sun of mass $M$. Geometrically the Runge-Lenz vector 
points towards the perihelion. Therefore its precession rate is just the 
precession rate of the perihelion \cite{13-E}.

However, we will use not the Runge-Lenz vector but its less known cousin,
the Hamilton vector \cite{8,9,10,11}
\begin{equation}
\vec{u}=\vec{v}-\frac{\alpha}{L}\,\vec{e}_\varphi,
\label{eq2}
\end{equation}
where $\varphi$ is the polar angle in the orbit plane. This very useful 
vector constant of motion of the Kepler problem was well known  in the past, 
but mysteriously disappeared from textbooks after the first decade of the 
twentieth century \cite{8,10,11,12}.

Of course, $\vec{A}$ and $\vec{u}$ are not independent constants of motion. 
The relation between them is
\begin{equation}
\vec{A}=\vec{u}\times\vec{L}.
\label{eq3}
\end{equation}
Remembering that the magnitude of the Runge-Lenz vector is $A=\alpha\,e$,
where $e$ is the eccentricity of the orbit \cite{5}, we get from (\ref{eq3})
the magnitude of the Hamilton vector
\begin{equation}
u=\frac{\alpha\,e}{L}.
\label{eq4}
\end{equation}
If the potential $U(r)$ contains a small central-force perturbation $V(r)$
besides the Coulomb binding potential,
$$U(r)=-\frac{\alpha}{r}+V(r),$$
the Hamilton vector (as well as the Runge-Lenz vector) ceases to be conserved
and begins to precess with the same rate as the Runge-Lenz vector, because
according to (\ref{eq3}) the two vectors are perpendicular.

To calculate the precession rate of the Hamilton vector we first find its
time derivative
\begin{equation}
\dot{\vec{u}}=-\frac{1}{\mu}\,\frac{dV(r)}{dr}\,\vec{e}_r,
\label{eq5}
\end{equation}
where $\mu=\frac{mM}{m+M}$ is the reduced mass. To get (\ref{eq5}), we have
used Newton's equation of motion for $\dot{\vec{v}}$ and the equation
$\dot{\vec{e}}_\varphi=-\dot{\varphi}\,\vec{e}_r$.

\noindent
Now the precession rate of the vector $\vec{u}$ can be found as \cite{6}
\begin{equation}
\vec{\omega}=\frac{\vec{u}\times\dot{\vec{u}}}{u^2}.
\label{eq6}
\end{equation}
Because of (\ref{eq2}), (\ref{eq5}) and (\ref{eq6}) only the tangential
component $r\dot{\varphi}$ of the velocity vector $\vec{v}=\dot{r}\vec{e}_r
+r\dot{\varphi}\,\vec{e}_\varphi$ contributes and we get
$$\vec{\omega}=\frac{1}{\mu u^2}\,\left (r\dot{\varphi}-\frac{\alpha}{L}
\right )\,\frac{dV(r)}{dr}\,\vec{e}_r\times\vec{e}_\varphi=
\frac{p}{\alpha e^2}\left (r\dot{\varphi}-\frac{\alpha}{L}\right )
\frac{dV(r)}{dr}\,\vec{k},$$
where $\vec{k}$ is the unit vector in the $z$-direction and
\begin{equation}
p=\frac{L^2}{\mu \alpha}
\label{eqp}
\end{equation}
is the semi-latus rectum of the unperturbed orbit.

Therefore, under the complete orbital cycle the Hamilton vector and hence 
the perihelion of the orbit revolves by the angle
\begin{equation}
\Delta \Theta_p=\int\limits_0^T\omega\,dt=\frac{p}{\alpha e^2}
\int\limits_0^{2\pi}\left (r-\frac{\alpha}{L\dot{\varphi}}\right )
\frac{dV(r)}{dr}\,d\varphi.
\label{eq7}
\end{equation}
However, $L=\mu r^2\dot{\varphi}$ and, therefore,
$$\frac{\alpha}{L\dot{\varphi}}=\frac{r^2}{p}.$$
Besides, to first order in the perturbation potential $V(r)$ we can use
the unperturbed orbit equation
$$\frac{p}{r}=1+e\cos{\varphi}$$
while integrating (\ref{eq7}) and get
\begin{equation}
\Delta \Theta_p=\frac{p^2}{\alpha e}\int\limits_0^{2\pi}\frac{\cos{\varphi}}
{(1+e\cos{\varphi})^2}\,\frac{dV(r)}{dr}\,d\varphi.
\label{eq8}
\end{equation}
If now we introduce the new integration variable $z=\cos{\varphi}$, 
equation (\ref{eq8}) transforms into
\begin{equation}
\Delta \Theta_p=-\frac{2p}{\alpha e^2}\int\limits_{-1}^1\frac{z}
{\sqrt{1-z^2}}\,\frac{dV\left (\frac{p}{1+ez}\right )}{dz}\,dz,
\label{eq9}
\end{equation}
and this is just equation (30) from \cite{1} up to the applied notations.

The extreme simplicity of this back-of-envelope derivation demonstrates 
clearly that the real backbone behind the Adkins and McDonnell perihelion
precession formula is the Hamilton's vector, the lost sparkling diamond
of introductory level mechanics.

After this work had been completed, we became aware of the paper \cite{13}, in 
which the author rediscovers the Hamilton vector and advocates essentially the 
same treatment of the perihelion precession problem as described in this note.
 ``Everything has been said before, but since nobody listens we have to keep 
going back and beginning all over again'' \cite{14}.

\section*{Acknowledgments}
An occasional  conversation with I.~B.~Khriplovich triggered our interest 
to this problem. The work of Z.K.S. is supported in part by grants
Sci.School-905.2006.2 and RFBR 06-02-16192-a.

\end{document}